\documentclass[12pt]{article}
\usepackage[margin=3cm]{geometry}
\usepackage{amsmath}
\usepackage{graphicx}

\begin{document}


\thispagestyle{empty}

\vspace*{0.3in}

\begin{center}
{\large \bf Quantum UV/IR Relations and Holographic Dark Energy from Entropic Force}

\vspace*{0.5in} {Miao Li$^{1}$ and Yi Wang$^{2}$}
\\[.3in]
{\em
$^1$ Kavli Institute for Theoretical Physics China,\\
Key Laboratory of Frontiers in Theoretical Physics,\\
Institute of Theoretical Physics, Chinese Academy of Sciences,\\
Beijing 100190, P.~R.~China\\
\vspace{.1in}
$^2$ Physics Department, McGill University,\\ Montreal, H3A2T8, Canada
\vspace{0.3in} }
\end{center}

\begin{center}
{\bf Abstract}
\end{center}
\noindent

We investigate the implications of the entropic force formalism proposed by Verlinde. We show that an UV/IR relation proposed by Cohen et al., as well as an uncertainty principle proposed by Hogan can be derived from the entropic force formalism. We show that applying the entropic force formalism to cosmology, there is an additional term in the Friedmann equation, which can be identified as holographic dark energy. We also propose an intuitive picture of holographic screen, which can be thought of as an improvement of Susskind's holographic screen.

\vfill

\newpage

\setcounter{page}{1}


\section{Introduction}

The investigation of black hole thermodynamics \cite{Bardeen:1973gs} implicate that there may be profound connections between gravity and thermodynamics. In \cite{Jacobson:1995ab}, Jacobson derived the Einstein equation from the thermodynamics near the horizon. The Einstein equation arises as an equation of state in the thermodynamical picture. Subsequently, this issue have been discussed by many authors. Especially, Padmanabhan discussed the equipartition rule and some other thermodynamics of gravity in \cite{Pad1}.

Recently, Verlinde \cite{Verlinde:2010hp} conjectured that gravity can be explained as an entropic force. In thermodynamics, if the number of states depends on position $\Delta x$, entropic force $F$ arises as thermodynamical conjugate of $\Delta x$. In this case, the first law of thermodynamics can be written as
\begin{equation}\label{entropyforce}
  F\Delta x=T\Delta S~.
\end{equation}

Inspired by Bekenstein's entropy bound, Verlinde postulated that when a test particle moves towards a holographic screen, the change of entropy on the holographic screen is proportional to the mass $m$ of the test particle, and the distance $\Delta x$ between the test particle and the screen:
\begin{equation}\label{assumption}
  \Delta S = 2\pi k_B \frac{mc}{\hbar}\Delta x~.
\end{equation}
To derive the entropic force hypothesis, Eq. \eqref{assumption} should hold at least when $\Delta x$ is smaller than or comparable with the Compton wavelength of the particle.

The temperature appears in Eq. \eqref{entropyforce} can be understood in two ways: one can relate temperature and acceleration using Unruh's rule
\begin{equation}\label{unruh}
  k_BT=\frac{1}{2\pi}\frac{\hbar a}{c}~,
\end{equation}
or relate temperature, energy and the number of used degrees of freedom using the equipartition rule
\begin{equation}\label{equip}
  E=\frac{1}{2}Nk_BT~.
\end{equation}

The role that the above equations play is as follows: Eq. \eqref{assumption} can be thought of as a basic assumption throughout Verlinde's work. Eq. \eqref{entropyforce} is an equation of force; Eq. \eqref{unruh} is an equation of acceleration, thus an equation of inertia; Eq. \eqref{equip} encodes the information of Newtonian gravity. Keeping these in mind, one can use Eq. \eqref{assumption} together with \eqref{entropyforce} and \eqref{unruh} to obtain $F=ma$. Newton's law of gravity $F=GMm/R^2$ can be obtained from Eqs. \eqref{assumption}, \eqref{entropyforce} and \eqref{equip} together
with a formula for $N$. And  using Eqs. \eqref{assumption}, \eqref{unruh} and \eqref{equip}, one can obtain a relation between entropy, used bits and Newtonian potential\footnote{One should note that the temperature $T$ in Eq. \eqref{unruh} and Eq. \eqref{equip} have different meaning. In Eq. \eqref{unruh}, the temperature is defined in the bulk. However, in Eq. \eqref{equip}, the temperature is defined on the holographic screen. To let these two temperatures equal is an additional assumption in Verlinde's paper.}:
\begin{equation}\label{master}
  \frac{S}{n}=-k_B\frac{\Phi}{2c^2}~.
\end{equation}
Eq. \eqref{master} has interesting physical implications. The ratio $-\Phi/(2c^2)$ takes value between 0 and 1. The implication is that the number of bits on the holographic screen which are used to dually describe the object in the bulk can be either equal to or larger than the entropy of the bulk object. In other words, Newtonian potential results in a ``coarse graining'' description of the bulk object on the holographic screen. This coarse graining is consistent with the picture of the AdS/CFT correspondence.

The entropic force conjecture of gravity can be applied to various aspects of gravitational physics. For example, this conjecture is applied to cosmology in \cite{Shu, Pad}. The implication for loop quantum gravity is discussed in \cite{Smolin}.

In this paper, we would like to discuss implications of Eq. \eqref{master}. In Section \ref{cohen}, we show that from the assumption of entropic force, one can derive two known holographic relations, namely, an UV/IR relation proposed by Cohen et al. \cite{Cohen:1998zx}, and an uncertainty principle proposed by Hogan \cite{Hogan:2007hc}.

In Section \ref{hde}, we discuss the implication of the entropic force for cosmology, especially dark energy. We find that the entropic force conjecture leads to Friedmann equation, which is consistent with the recent studies \cite{Shu, Pad}. However, we note that one additional term in the Friedmann equation arises, which can be identified to holographic dark energy. \footnote{
Similar topics have also been discussed in \cite{Lee:2007zq} and \cite{Koelman}. The differences between these previous works and the present Letter are as follows: In \cite{Lee:2007zq}, holographic dark energy is derived using information theory, which is similar to, but not the same as Verlinde's entropic force formalism. In \cite{Koelman}, dark energy is derived from Verlinde's entropic force formalism. However, there the cosmic horizon is associated with the scale factor $a$. Thus the result is not canonical. }

In Section \ref{screen}, we visualize the discussion of Section \ref{cohen} by constructing an improved holographic screen. We shall show that the intuitive holographic screen proposed by Susskind \cite{Susskind:1994vu} has tension with Eq. \eqref{master}, and the entropic force conjecture leads to a solution to this problem.

For simplicity, we shall use natural units $\hbar=c=k_B=1$ in the remainder of the Letter.

\section{Holographic relations from entropic force} \label{cohen}

Since the discovery of holography, UV/IR correspondence has become an important concept in physics concerning gravity. It is conjectured that when gravity is considered, the UV and IR cutoffs of an effective field theory should be related. When an IR cutoff of an effective field theory is chosen, an UV cutoff arises according to this IR cutoff. Thus one can write the UV cutoff as a function of the IR cutoff as
\begin{equation}\label{eq:ansatz}
  L_{UV}=f(L_{IR})~.
\end{equation}

To apply Eq. \eqref{master}, we consider how information on the horizon of a Schwarzchild black hole is coarse gained on a holographic screen. We use $L$ to denote the distance between the black hole and the holographic screen. Thus the amount of coarse graining is
\begin{equation}\label{eq:rat1}
  \frac{\Delta A_h}{\Delta A_s}=-\frac{\Phi}{2} = \frac{L_p^2 M}{2L} ~,
\end{equation}
where $L_p\equiv \sqrt{G}$ is the Planck length. Here we use $\Delta A_h$ and $\Delta A_s$ to denote the fundamental area elements on the black hole horizon and on the holographic screen respectively. By ``fundamental'', we mean the smallest area that can be treated semiclassically. Thus the fundamental area elements are related to the UV cutoffs of the theory as
\begin{equation}\label{eq:rat2}
  \Delta A_h=f^2(\alpha r_h)= f^2(2\alpha L_p^2 M)~,\qquad
  \Delta A_s=f^2(L)~,
\end{equation}
where $r_h$ is the Schwarzchild radius. Note that the natural IR cutoff for the holographic screen is given by the distance $L$. The IR cutoff for the black hole horizon should be proportional to $r_h$, and we use a constant factor $\alpha$ to denote the numerical coefficient $L_{IR}/r_h$.

Comparing Eqs. \eqref{eq:rat1} and \eqref{eq:rat2}, we have
\begin{equation}\label{eq:comp}
  \frac{\Delta A_h}{\Delta A_s}= \frac{L_p^2 M}{2L}=\frac{ f^2(2\alpha L_p^2 M)}{f^2(L)}~.
\end{equation}

Note that Eq. \eqref{eq:comp} should be valid for arbitrary $L$ and $M$ . Thus we have
\begin{equation}\label{eq:sol}
  f(L_{IR})=\sqrt{\beta L_p L_{IR}}~,\qquad \alpha=1/4~,
\end{equation}
where $\beta$ is a dimensionless numerical constant, which canceled out in Eq. \eqref{master}, thus remains not determined. Note that $\beta$ cannot depend on $L$ or $M$. Because $L$ and $M$ can vary for different holographic screens and different black holes, while Eq. \eqref{eq:ansatz} should be a general law. The natural value of $\beta$ is order one.

Insert Eq. \eqref{eq:sol} into Eq. \eqref{eq:ansatz}, we have
\begin{equation}\label{eq:uvirL}
  L_{UV}=\sqrt{\beta L_p L_{IR}}~.
\end{equation}

In terms of the UV/IR cutoffs of energy scales, the above equation takes the form
\begin{equation}\label{eq:uvir}
  \Lambda_{UV}^2=\sqrt{8\pi}\beta M_p \Lambda_{IR}~,
\end{equation}
where $M_p\equiv 1/\sqrt{8\pi G}$ is the reduced Planck mass. This equation is the equation proposed in Cohen's et al. paper \cite{Cohen:1998zx}, which is originally obtained by requiring that the vacuum energy inside a volume characterized by $L_{IR}$ does not exceed the energy of a black hole in this volume.

One outstanding application of Cohen's et al. UV/IR relation is holographic dark energy \cite{Li:2004rb}, which provides a solution of the cosmological constant problem and can fit the current data very well. In the model of holographic dark energy, the energy density of vacuum energy takes the form
\begin{equation}\label{hde}
  \rho_{\Lambda}=3c^2 M_p^2 R_h^{-2}~.
\end{equation}
The form of holographic dark energy can be obtained by noticing the fact that $\rho_\Lambda\sim \Lambda_{UV}^4$, and choosing the IR cutoff as the future event horizon.

Before to proceed, we would like to mention that the UV/IR relation \eqref{eq:uvirL} can also be obtained in another way. We consider the case that a fundamental area element of a black horizon is represented on two different holographic screens. The distance between the black hole and these two holographic screens are denoted by $L_1$ and $L_2$ respectively. The fundamental area element of these two holographic screens $\Delta A_1$, $\Delta A_2$ satisfy
\begin{equation}\label{eq:alteruvir}
  \frac{\Delta A_1}{\Delta A_2}=\frac{f^2(L_1)}{f^2(L_2)}=\frac{\Phi_2}{\Phi_1}=\frac{L_1}{L_2}~,
\end{equation}
where $\Phi_1$ and $\Phi_2$ are Newtonian potentials on the holographic screen. As $L_1$ and $L_2$ can be chosen arbitrarily, we have $L_{UV}=f(L_{IR})\propto \sqrt{L_{IR}}$. This recovers Eq. \eqref{eq:uvirL}.  Note that suppose there are $n_i$ fundamental area elements on holographic screen $i$ $(i=1,2)$. Then $n_i=A_i/\Delta A_i$, where $A_i$ is proportional to all the degree of freedom used to describe the massive object. Here our derivation also works for a mass not necessarily a black hole.

\begin{figure}
  \centering
  \includegraphics[width=0.6\textwidth]{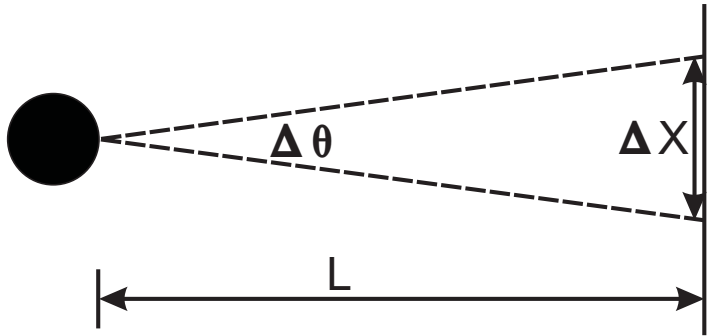}
  \caption{\label{fig:hogan} When a light ray travels a distance $L$, the perpendicular direction obtains an uncertainty $\Delta X$. Equivalently, the direction of the light ray obtains an uncertainty $\Delta \theta$.}
\end{figure}

As another application of Eq. \eqref{master}, in the remainder of this section, we shall derive a uncertainty relation of quantum gravity. We consider when a light ray travels a distance $L$, how precisely the direction that the light ray travels can be determined. To be more explicit, we assume when a light ray travels a distance $L$, the direction of the light ray has an uncertainty $\Delta\theta\equiv g(L)$. This uncertainty is illustrated in Fig. \ref{fig:hogan}.

To determine the function $g(L)$, we consider the same setup in deriving \eqref{eq:alteruvir}. Two holographic screens are set up with distance $L_1$ and $L_2$ respectively from a gravitational source, which produces a Schwarzchild type Newtonian potential. The level of fuzziness of a light ray which travels from the gravitational source to the holographic screens can be thought of as a measure of coarse graining. This is because, if the light rays were not fuzzy in the perpendicular direction, one could use light rays to communicate information between the gravitational source and the holographic screens to make correspondence for every bit of information, such that no coarse graining is necessary. To equal fuzziness and amount of coarse graining, we have
\begin{equation}
  \frac{\Delta A_1}{\Delta A_2}=\frac{g^2(L_1)L_1^2}{g^2(L_2)L_2^2}=\frac{\Phi_2}{\Phi_1}=\frac{L_1}{L_2}~.
\end{equation}
The above equation holds for general $L_1$ and $L_2$. Thus we have
\begin{equation}\label{eq:hogan}
  \Delta\theta^2=\gamma \frac{L_p}{L}~,
\end{equation}
where $\gamma$ is a dimensionless constant. Eq. \eqref{eq:hogan} recovers the holographic uncertainty relation proposed by Hogan \cite{Hogan:2007hc}. This relation can be also expressed by a uncertainty of distance $\Delta X = \Delta\theta L$ in the direction perpendicular to $L$. In terms of $\Delta X$, the uncertainty principle takes the form
\begin{equation}
  \Delta X^2=\gamma L_p L~.
\end{equation}
Originally, Hogan's uncertainty principle was proposed by making use of the diffraction equation of a Planck wavelength light ray to get the fuzziness. Alternatively, Hogan's uncertainty principle can be also understood as: at Planck scale spacetime foam, when a light ray travels a distance $L_p$, the uncertainty in the perpendicular direction is also of order $L_p$. When the light ray travels a distance $L$, which has $L/L_p$ intervals of Planck distance, the random walk in the perpendicular direction accumulates to be $\Delta X \sim L_p\times \sqrt{L/L_p}=\sqrt{L L_p}$. This uncertainty principle can also be related to the UV/IR relation Eq. \eqref{eq:uvirL} by identifying $\Delta X$ to be the UV cutoff, and $L$ to be the IR cutoff.

Finally, one should note that the derivation in this section is reversible. If we start from Cohen's et al. UV/IR relation or start from Hogan's uncertainty principle, one can derive Verlinde's formula \eqref{master}. And from Eq. \eqref{master}, one can arrive at Eq. \eqref{assumption}, which is the basic assumption of Verlinde's paper.

\section{Implications for cosmology and holographic dark energy} \label{hde}

As discussed in the previous section, Cohen's et al. UV/IR relation arises in the framework of entropic force. Following the arguments in \cite{Li:2004rb}, this UV/IR relation leads to a model of holographic dark energy. In this section, we shall show that the previous argument of dark energy can also be obtained from the analysis of the future event horizon and a holographic screen describing the observable universe.

\begin{figure}
  \centering
  \includegraphics[width=0.6\textwidth]{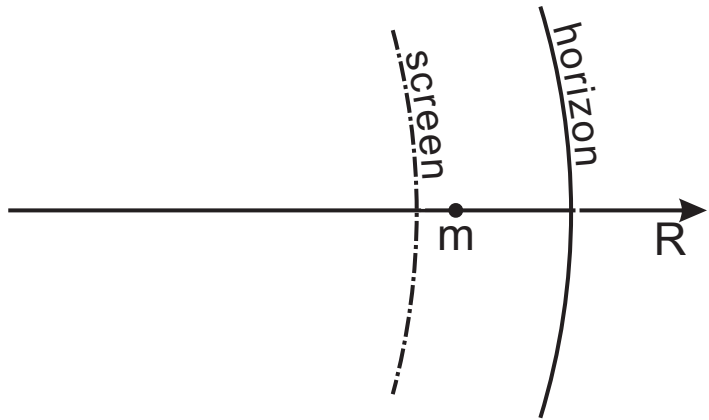}
  \caption{\label{fig:hde} To investigate the role that future event horizon plays in cosmology, we consider the setup as illustrated in the figure. The point $m$ denotes a test particle located near the holographic screen.}
\end{figure}

In spite of ordinary matter, our universe also very probably has a future event horizon. Consider a test particle which lies slightly outside a holographic screen, but ``inside'' the future event horizon of the universe. We assume the distance between the test particle and the future event horizon (seen from an observer in the center of the observable universe) be much larger than a Planck length, so that Newtonian approximation is valid. On the other hand, this distance should be also smaller than the size of the observational universe (in order that the holographic screen is also of cosmological size), so that one can investigate cosmological sequences. We illustrate this setup in Fig. \ref{fig:hde}.

The future event horizon of the universe has Gibbons-Hawking radiation $T_h\sim H\sim 1/R_h$, as well as degree of freedom $N_h\sim R_h^2/G$.  As noticed by Verlinde, the energy of the future event horizon, which is seen on the holographic screen, takes the form
\footnote{As a check, one can consider the case of a black hole. Consider a screen which lies far away from the black hole. We denote the Schwarzchild radius by $R_s$. The Hawking temperature of the black hole seen on the holographic screen is $T\sim 1/R_s$, the degree of freedom of the black hole horizon is $N\sim R_s^2/G$. Thus seen from the holographic screen, the black hole has energy $E_{BH} \sim R_s/G$, which is indeed the mass of the black hole. }
\begin{equation}\label{eq:rhenergy}
  E_h\sim N_hT_h\sim R_h/G~.
\end{equation}

We denote the energy density of matter components (for example, dust and radiation) by $\rho_m$. The energy represented on the holographic screen is
\begin{equation}\label{eq:eh}
  E=\frac{4\pi R^3}{3}\rho_m ~,
\end{equation}
where $R$ is the physical radial coordinate of the test particle.

Following Verlinde's logic, using Eqs. \eqref{entropyforce}, \eqref{assumption} and \eqref{equip}, we find that the energy on the holographic screen induces a force on the test particle toward it, while the event horizon induces another force pointing outward along R. Considered these facts, the force experienced by the test particle takes the form \footnote{A black hole horizon and a cosmic horizon does not make much difference, except that for the cosmic horizon, objects exits the horizon from inside to outside. Thus Verlinde's entropic derivation on the force should also apply for the cosmic horizon: as there is an entropy change, there should be a force. This is different from directly applying Newtonian gravity, where there should be no force outside the spherical shell from cancelations between different parts of the shell.}
\begin{equation}
  F\sim \frac{mR_h}{R^2}-4\pi G mR\rho_m~£¬
\end{equation}
where the temperature $T$ is defined in Eq. \eqref{equip} by the equipartition rule. Note that we have also assumed the degree of freedom on the holographic screen has energy $T_h$. However, this is not a problem because $T_h\sim T$ in our setup. Note that the term $4\pi G R \rho_m$ corresponds to the Newtonian gravity from the matter components. However, a new term $mR_h/R^2 \equiv F_h$ arises because of the existence of the future event horizon. We find an attractive force and a repulsive force, the latter is due to
dark energy.

The potential energy for the test particle arising from the effect of the future event horizon is
\begin{align}\label{eq:vhde}
  V_h \sim -\frac{R_h m}{R} = -\frac{c^2mR^2}{2 R_h^2}~,
\end{align}
where $c$ is a numerical constant. Note that at the equal sign, we have used the fact that from our setup, $R\sim R_h$. Thus LHS and RHS are almost equal\footnote{Note that as we have $T\sim T_h$ and $R\sim R_h$, if we slightly change our derivation, we might have obtained equations for $F_h$ up to a factor $R_h/R$. But as our result is derived up to a multiplying numerical factor, the factor $R_h/R$ can be ignored even when one integrates $F_h$ to obtain $V_h$. One exception is that if $F\sim m/R$, there will be a logarithm $\log(R_h/R)$ in $V_h$. But again, this logarithm is order one, and does not change the main result here.}. Note that the time derivative of $R$ and $R_h$ are different. However, as one can see from the remainder of this section, we shall not take derivative with respective to $R$ or $R_h$ any more, thus the replacement between $R_h$ and $R$ is allowed here. As we shall show, this additional gravitational potential for the test particle can be identified as holographic dark energy.

To see the implication of Eq. \eqref{eq:vhde}, we derive the Friedmann equation. As discussed by Verlinde, the entropic force conjecture leads to the Newtonian gravity. Applying to cosmology, there are several methods to derive Friedmann equation. For example, one can use the Tolman-Komar energy \cite{Pad}, or use some properties of apparent horizon \cite{Shu}. Here we shall review a simpler pedagogical way to derive Friedmann equation from Newtonian gravity:

The total energy of the test particle can be written as $E=m\dot{R}^2/2+V$~, where
\begin{equation}\label{eq:vtotal}
  V\equiv V_{m}+V_{h} = -\frac{4\pi G}{3}m\rho_m R^2-\frac{c^2mR^2}{2R_h^2}~.
\end{equation}

Note that the total energy $E$ of the test particle should be a constant. We write the physical radius $R=ar$, where $a$ is the scale factor and $r$ is the comoving coordinate of the particle, which is by definition a constant. Divide Eq. \eqref{eq:vtotal} by $mR^2/2$, we have the Friedmann equation
\begin{equation}\label{eq:fried}
  3M_p^2 H^2=\rho_m + \rho_k+ 3c^2M_p^2R_h^{-2}~,
\end{equation}
where $\rho_k\equiv 6M_p^2 E/(mR^2)$ is the effective energy density for the spatial curvature of the universe. Thus holographic dark energy arises as a consequence of force experienced by a test particle near a cosmological size holographic screen with the presence of future event horizon.

Finally, we would emphasize that we have obtained the energy density of holographic dark energy, but by far not the equation of state. There are two reasons for this. One reason is that here we have used the Newtonian approximation of gravity. If we have not started from the conservation of energy, but rather from the equation $F=m\ddot R$ directly, we shall not obtain the correct Friedmann equation without replacing $M\rightarrow M+3M'$ \cite{Pad}. Thus one cannot confidently take time derivative of Eq. \eqref{eq:fried} in the framework of Newtonian gravity. Another reason is that we have made replacements between $R$ and $R_h$ in Eq. \eqref{eq:vhde}. This shall not allow us to further take time derivative to \eqref{eq:fried}.

To obtain the equation of state for holographic dark energy, we should note that the energy density of holographic dark energy should only depend on the size of the future event horizon, but not depend on the scale factor of the universe. A supporting argument is that in a de Sitter universe, due to the scale invariance of the de Sitter phase, the energy density of holographic dark energy should not dependent on the scale factor. This is different with the term $\rho_k$, which is absent in the pure de Sitter phase. Similarly, it is natural that in a quasi-de Sitter phase, which can describe the current universe, the energy density of holographic dark energy should also not depend on the scale factor. Thus the energy density of holographic dark energy should be $\rho_{\Lambda}=3c^2M_p^2R_h^{-2}$, instead of something like $3c^2M_p^2R_h^{-1}R^{-1}$, $3c^2M_p^2R^{-2}$, etc.

The above argument can also obtained in an simper way: The scale factor is not a canonical quantity, namely when we rescale spatial coordinates $r$, the scale factor also gets rescaled, thus the dependence of the dark energy density on the scale factor cannot be simply a power form.

Now we can take time derivative to the holographic dark energy density, resulting in
\begin{equation}\label{eq:eoshde}
  \dot\rho_{\Lambda}=-2\rho_{\Lambda}(H-R_h^{-1})~.
\end{equation}
Eqs. \eqref{eq:fried} and \eqref{eq:eoshde} fully determines the dynamics of holographic dark energy.

\section{Towards an improved holographic screen} \label{screen}

In this section, we discuss the tension between Susskind's holographic screen \cite{Susskind:1994vu} and Verlinde's entropic force conjecture. Based on this, we propose an improved intuitive picture of holographic screens.

\begin{figure}
  \centering
  \includegraphics[width=0.4\textwidth]{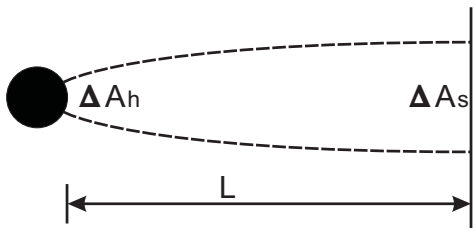}
  \caption{\label{fig:susskind}Susskind's holographic screen.}
\end{figure}

In \cite{Susskind:1994vu}, Susskind proposed an intuitive picture of holographic screen. One can think of a holographic screen as a spacelike plane. The image of a black hole on the screen is defined by the intersection of the holographic screen with the set of light rays that start from the stretched horizon of black hole, and end on the screen at right angle. This picture is illustrated in Fig. \ref{fig:susskind}.

We would like to examine the holographic feature of Susskind's screen in Verlinde's picture. Consider a black hole with fixed entropy $S$. The bits used on the holographic screen to describe the black hole can be calculated using Eq. \eqref{master} as
\begin{equation}
  n=-\frac{2c^2}{\Phi}\frac{S}{k_B}.
\end{equation}
As one moves the holographic screen farther away from the black hole, which corresponds to coarse graining of information, the Newtonian potential decreases. Thus one need to use more bits on the holographic screen to describe the black hole. Qualitatively, this agrees with Susskind's screen.

However, there is tension between Eq. \eqref{master} and Susskind's screen when one performs a quantitative analysis. This is because, if one moves the screen to infinity, corresponding to a vanishing of gravitational potential, then from Eq. \eqref{master}, $n$ diverges. In other words, the image on the screen will be infinitely large.

On the other hand, the image has finite area from Susskind's definition of holographic screen, even if the screen is infinitely far away. This is because a light ray with impact parameter $b>3\sqrt{3}GM$ will not hit the black hole. Thus when the screen is infinitely far away, the radius of the image is $3\sqrt{3}GM$. \footnote{This image includes the ``primary screen map'' and other images, defined in \cite{Corley:1996qh}. Multiple images are possible here because light rays can orbit the black hole. Thus the light rays can be classified by number of cycles it goes around the black hole before hitting the holographic screen. The primary screen map is a one-on-one map from the black hole horizon to the holographic screen, which is in this case constructed by light rays that do not orbit the black hole. The primary screen map is smaller than $3\sqrt{3}GM$, which is also finite.}

This inconsistency leads to two possibilities: either Verlinde's entropic force hypothesis is not valid, or one need to improve the picture of holographic screen. The latter possibility is more likely to be true. This is because Eq. \eqref{master} is also supported by AdS/CFT, which represents a more modern point of view of holography.

\begin{figure}
  \centering
  \includegraphics[width=0.4\textwidth]{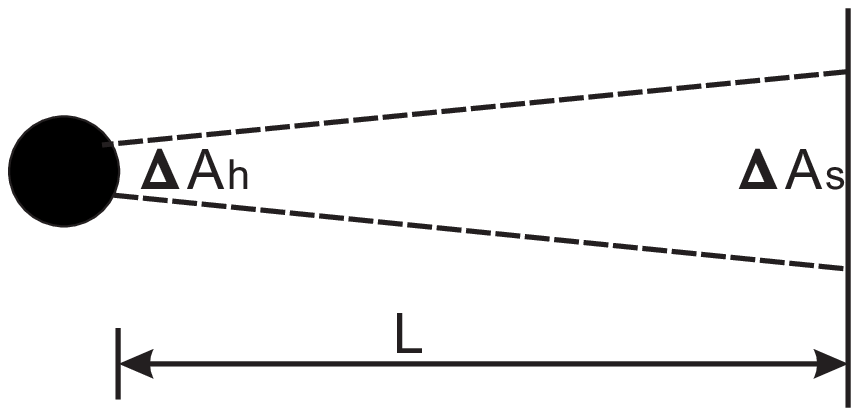}
  \hspace{1cm}
  \includegraphics[width=0.4\textwidth]{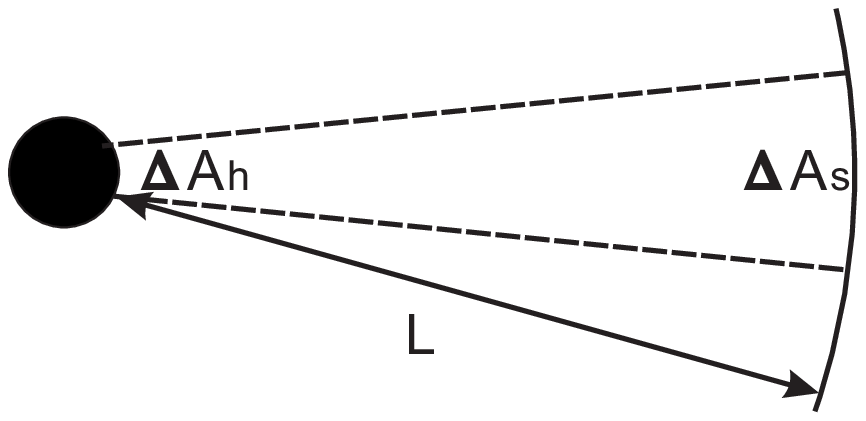}
  \caption{\label{fig:newscreen} Improved holographic screens. In the left figure, we illustrate the direct generalization of Susskind's holographic screen, which is planar. In the right figure, as Verlinde points out, we use an equipotential surface as a more natural choice of holographic screen.}
\end{figure}

Inspired by Hogan's uncertainty principle with the presence of gravity, we propose that the image of a fundamental region of black hole horizon should be presented by the fuzziness of a light ray travels from the stretched horizon to the holographic screen. This picture is illustrated in Fig. \ref{fig:newscreen}. Note that we could either construct holographic screens as planes, which are direct improvements of Susskind's holographic screen, or as spheres, which coincides better with Verlinde's view that equipotential surfaces are natural holographic screens.

\section{Conclusion and discussion}

We showed that Verlinde's entropic force formalism can be used to derive Cohen's et al. holographic UV/IR relation and Hogan's holographic uncertainty principle. The new derivation provides an improved theoretical background for these relations.

We showed that a component of dark energy arises from the entropic force formalism. This dark energy component can be identified with holographic dark energy. We are able to derive the energy density of holographic dark energy from the entropic force formalism. The equation of state for holographic dark energy is not obtained directly from this derivation. However, we can fix the equation of state of holographic dark energy by comparing our current universe with a pure de Sitter universe. The resulting equation of state also turns out to be the same as holographic dark energy.

We showed that Susskind's holographic screen is not consistent with Verlinde's entropic force formalism. Inspired by Hogan's uncertainty principle, we developed an improved intuitive picture of holographic screens. Despite of the improvement mentioned in the Letter, one should also keep in mind that holographic screens are only illustrations
of the holographic map in a loose sense. The precise correspondence of the bulk and the holographic screen requires more understanding of the dynamics of holography.

\section*{Acknowledgments}
YW would like to thank Alejandra Castro, Robert Brandenberger, Andrew Frey, Chunshan Lin, Alex Maloney, Omid Saremi, Marcus Tassler and Bret Underwood for discussion on the issue of entropic force. The research of LM was supported by a NSFC grant No.10535060/A050207, a NSFC grant No.10975172, a NSFC group grant No.10821504 and Ministry of
Science and Technology 973 program under grant No.2007CB815401.
YW was supported by NSERC and an IPP postdoctoral fellowship.

\end{document}